\documentclass[12pt, preprint]{aastex}
\begin{document}

\title{The Surface of 2003 $\rm EL_{61}$ in the Near Infrared}

\author{Chadwick A. Trujillo}
\affil{Gemini Observatory, Northern Operations Center, 670 N. A'ohoku
  Place, Hilo, Hawaii 96720}
\email{trujillo@gemini.edu}
\and
\author{Michael E. Brown}
\affil{California Institute of Technology, Division of Geological and
Planetary Sciences, MS 150-21, Pasadena, California 91125}
\email{mbrown@caltech.edu}
\and
\author{Kristina M. Barkume}
\affil{California Institute of Technology, Division of Geological and
Planetary Sciences, MS 150-21, Pasadena, California 91125}
\email{barkume@gps.caltech.edu}
\and
\author{Emily L. Schaller}
\affil{California Institute of Technology, Division of Geological and
Planetary Sciences, MS 150-21, Pasadena, California 91125}
\email{emily@gps.caltech.edu}
\and
\author{David L. Rabinowitz}
\affil{Yale Center for Astronomy and Astrophysics, Physics Department,
  Yale University, P.O. Box 208121, New Haven, Connecticut 06520-8121}
\email{david.rabinowitz@yale.edu}

\author{\large Accepted to be published in the Astrophysical Journal
  Jan 20, 2007, v655}

\begin{abstract}
We report the detection of crystalline water ice on the surface of
2003 $\rm EL_{61}$.  Reflectance spectra were collected from Gemini
North telescope from 1.0 to 2.4 \micron\/ wavelength range, and from
the Keck telescope across the 1.4 to 2.4 \micron\/ wavelength range.
The signature of crystalline water ice is obvious in all data
collected.  Like the surfaces of many outer solar system bodies, the
surface of 2003 $\rm EL_{61}$ is rich in crystalline water ice, which
is energetically less favored than amorphous water ice at low
temperatures, suggesting that resurfacing processes may be taking
place.  The near infrared color of the object is much bluer than a
pure water ice model.  Adding a near infrared blue component such as
hydrogen cyanide or phyllosilicate clays improves the fit
considerably, with hydrogen cyanide providing the greatest
improvement.  The addition of hydrated tholins and bitumens also
improves the fit but is inconsistent with the neutral $V-J$
reflectance of 2003 $\rm EL_{61}$.  A small decrease in reflectance
beyond 2.3 \micron\/ may be attributable to cyanide salts.  Overall,
the reflected light from 2003 $\rm EL_{61}$ is best fit by a model of
2/3 to 4/5 pure crystalline water ice and 1/3 to 1/5 near infrared
blue component such as hydrogen cyanide or kaolinite.  The surface of
2003 $\rm EL_{61}$ is unlikely to be covered by significant amounts of
dark material such as carbon black, as our pure ice models reproduce
published albedo estimates derived from the spin state of 2003 $\rm
EL_{61}$.
\end{abstract}


\section{Introduction}
\label{intro}

The Kuiper belt objects (KBOs) have been known for a decade to have
the most diverse surfaces of any minor planet population in terms of
global color at visible wavelengths \citep{1996AJ....112.2310L}.  The
carrier of this color has been tentatively identified as a red
tholin-like compound \citep{2001AJ....122.2099J}.  The situation is
much different in the near-infrared, where signatures of the most
primitive solar system volatiles can be seen.  Although many attempts
have been made to detect ices on KBOs in the near-infrared, very few
to date have been successful.

Barring the decades of study of Pluto, Charon and Triton (each of
which could share origins with the KBOs), detections of simple organic
ices on KBOs have been relatively few.  Indisputable methane ice
detections have been presented for 2003 $\rm UB_{313}$ and 2005 $\rm
FY_{9}$ \citep{2006btr,2005DPS....37.5211B,2006A&A...445L..35L}.  The
KBO Sedna has also been reported to have a Triton-like spectrum,
largely dominated by methane ice \citep{2005A&A...439L...1B}.  The
following KBOs have all shown signs of water ice: (19308) 1996 $\rm
TO_{66}$ \citep{1999ApJ...519L.101B}, (26375) 1999 $\rm DE_{9}$
\citep{2001AJ....122.2099J} and (90482) Orcus
\citep{2005ApJ...627.1057T}.  The object (50000) Quaoar has been the
only KBO so far to show signs of crystalline water ice
\citep{2004Natur.432..731J}.  This finding is significant in that
crystalline water ice is unstable on 10 Myr timescales.  However,
crystalline water ice has been seen on many other outer solar system
bodies such as Charon and the giant planet moons, which argues for a
common resurfacing process throughout the solar system.

In this work, we report the discovery of large amounts of crystalline
water ice on 2003 $\rm EL_{61}$ from observations obtained at the
Gemini North telescope and Keck Observatory.  The object 2003 $\rm
EL_{61}$ is a unique object in dynamical terms.  It is the only
known ternary or higher system in the Kuiper belt \citep[excepting
Pluto,][]{2005IAUC.8625....1W}, with the brighter moon's orbit reported
in \cite{2005ApJ...632L..45B} and the fainter moon
reported in \cite{2006betal}.  Not only is 2003 $\rm EL_{61}$ a
ternary system, but it is the most rapid large ($> 100$ km) rotator in
the solar system \citep{2005DPS....37.5612R}.  The extreme rotational
state of 2003 $\rm EL_{61}$ combined with the measured orbit of its
brighter moon yields a surprisingly large amount of information about
the physical parameters of the system including limits on its shape,
density and albedo \citep{2006retal}.  With the discovery of
crystalline water ice on the surface of 2003 $\rm EL_{61}$, we can now
place constraints on the surface fraction of water ice on the body as
well as other components.

\section{Observations}

Observations of 2003 $\rm EL_{61}$ were collected at the Gemini North
and the Keck telescopes on Mauna Kea, Hawaii.  The Gemini data were
collected in queue mode on 4 separate UT dates, as detailed in
Table~\ref{observations}.  The Keck observations were collected in
classical (i.e. full night) mode on UT 2005 Apr 26 and 27.  Total
integration times for Gemini were 4.0 hours in $K$-band, 1.3 hours in
$H$-band and 1.7 hours in $J$-band using NIRI
\citep{2003PASP..115.1388H}.  From Keck, total integration times were
3.1 hours in a single $HK$ grism setting using NIRC \citep{nirc}.
Some of the data were taken through cirrus, however, removal of the
cirrus data did not affect results, so they were included in this
work.  A telluric standard star was observed on each night with
similar airmass to the science observations.  On the two photometric
Gemini nights, photometry data for 2003 $\rm EL_{61}$ and UKIRT faint
standards were collected.

\section{Data Reduction}

\subsection{Photometry}

On the two photometric Gemini nights, synthetic aperture photometry
was used to measure the broad-band colors of 2003 $\rm EL_{61}$.  A
large aperture of 4.7 arcseconds diameter (40 pixels) was selected due
to the poor seeing (0.9 arcseconds full width at half maximum in $H$).
This chosen aperture encompassed the secondary, which is up to 1
arcsecond away from and 3.3 magnitudes fainter than the primary.  The
primary has a known visible lightcurve with peak-to-peak amplitude
$0.28 \pm 0.04$ magnitudes and a period of $3.9154 \pm 0.0002$ hours
in the visible without color dependence \citep{2006retal}.  Flux
measurements for $J$, $H$, and $K$ were made during UT 2005 Mar 02
10:34 -- 11:08 and also on UT 2005 Apr 25 07:25 -- 07:44.  Flux
measurements were compared to the Mauna Kea JHK system UKIRT standard
stars FS23 and FS131 on the two respective nights
\citep{2002PASP..114..169S,2002PASP..114..180T,letalprep}.
Near-infrared colors were consistent between the two nights with $J-H
= -0.044 \pm 0.037 $ and $H-K = -0.111 \pm 0.048$.  These
near-infrared colors were used to later scale the Gemini $J$, $H$ and
$K$ grism settings to a common relative reflectance scale.  This scale
was consistent with the lower resolution Keck spectrum which collected
$H$ and $K$ using a single grism setting.

The individual $J$, $H$ and $K$ magnitudes were all brighter on UT
2005 Apr 25 by an average of $0.149 \pm 0.042$ magnitudes, which is
consistent with the visible lightcurve measured by \cite{2006retal}.
Thus, it is likely that the visible and near infrared lightcurves have
the same shape and phase information, just as the $B$, $V$, $R$, and
$I$ colors do, as would be expected for a shape-dominated lightcurve.
The two near infrared photometry epochs (UT 2005 Mar 02 10:51 and UT
2005 Apr 25 07:35) correspond to the lightcurve phases 0.834 and
0.016, which occur at $-0.06$ and $0.07$ magnitudes from the mean
visible magnitude.  Using the assumption that the near-infrared
lightcurve and visible lightcurves match in phase and magnitude, we
find the following for the mean near-infrared magnitudes for 2003 $\rm
EL_{61}$: $J = 16.518 \pm 0.020$, $H = 16.561 \pm 0.031$, $K = 16.672
\pm 0.037$.  These numbers are uncorrected for the solar phase angle
because solar phase corrections are unmeasured in the near infrared
and likely differ from the visible.  However, the solar phase angles
for the two epochs are similar (0.71 degrees and 0.65 degrees,
respectively), so the above magnitudes should be valid for the mean
solar phase angle of 0.68 degrees.  From \cite{2006retal}, we
determine the mean visible magnitude to be $V = 17.569 \pm 0.019$ at
0.68 degrees solar phase and heliocentric and geocentric distances of
$R = 51.254$ AU and $\Delta = 50.461$ AU, corresponding to the
midpoint between our near infrared photometric epochs.  Thus we
produce the following visible to near-infrared color: $V-J = 1.051 \pm
0.020$, which is consistent with solar $V-J$ color
\citep{2001AJ....122.2099J}.

\subsection{Spectroscopy}

Both the Gemini and Keck spectral data were processed using standard
near infrared techniques.  First, spectra dithered along the slit were
pairwise sky subtracted to remove detector bias and dark current as
well as produce a first order removal of sky lines.  These images were
divided by a composite flat of the slit observed through the grism.
The resultant image pairs were cleaned of residual bad pixels and
cosmic rays.  Images were rectified such that the wavelength and
spatial axes were orthogonal prior to spectral extraction.  Source
spectra were extracted from each dither position along with adjacent
sky regions.  A composite residual sky spectrum was subtracted from a
composite source spectrum for each night of data.  Telluric G2V star
spectra were put through the same processing pipeline prior to being
divided into the source spectra for telluric correction to produce
relative reflectance.  Data from each night were then combined into a
single science spectrum.

\begin{figure}
\plotone{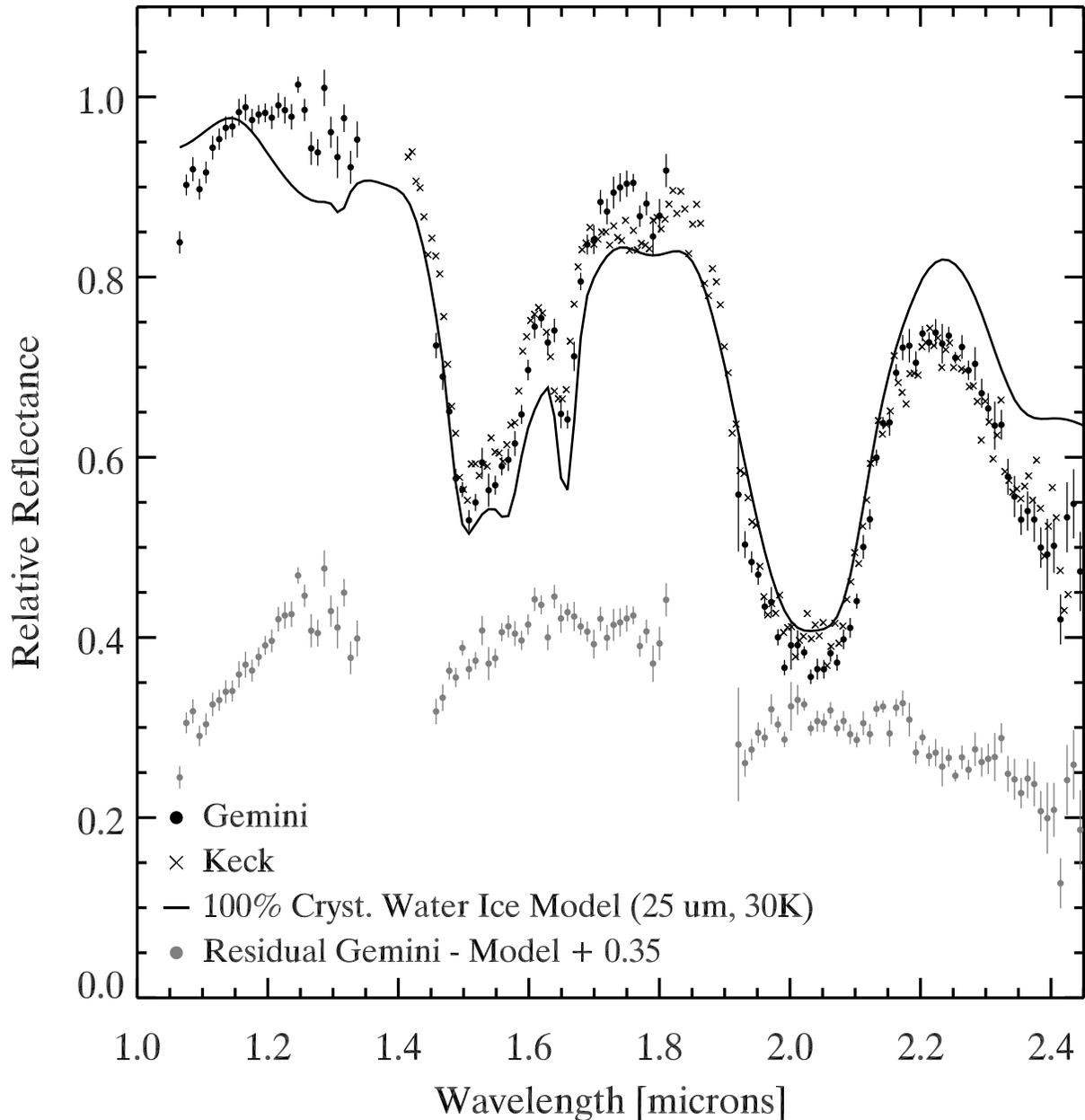} \figcaption{Relative reflectance spectrum
  of 2003 $\rm EL_{61}$ from Gemini and Keck (black filled circles and
  x's) normalized to the model.  Overplotted is our best fit 100\%
  pure crystalline water ice model at its true geometric albedo.  The
  spectrum - model residuals (gray filled circles) illustrate that an
  additional blue component is needed to reproduce the near-infrared
  colors of 2003 $\rm EL_{61}$, particularly in the $H$ to $K$ region.
  See text for full details of the model.
\label{el61-waterice}}
\end{figure}

Finally, the spectral resolution of the Gemini data was reduced to
match the Keck NIRC data after a search for narrow-band features
resulted in no significant detections.  The Gemini spectrum was
converted to reflectance by using the measured $J-H$ and $H-K$ colors
of 2003 $\rm EL_{61}$ to scale the 3 separate $J$, $H$ and $K$ grism
settings of NIRI.  Since no photometric data were collected at the
Keck telescope, we scaled the single $HK$ grism setting for Keck NIRC
to the Gemini spectrum.  The color-corrected Gemini spectrum in the
$H$ and $K$ regions was consistent with the single Keck $HK$ grism
setting, so we believe that both spectra are accurate in terms of
reflectance.  The data from the two telescopes can be found in
Figure~\ref{el61-waterice}.  We found no significant evidence for
rotational variation of spectral features in our data.  There was some
variation in the 1.7 \micron\/ to 1.8 \micron\/ range between the March
and April Gemini data ($\sim 10$\%) as well as the between the Gemini
data and the Keck data ($\sim 5\%$, seen in
Figure~\ref{el61-waterice}).  We note that this region has many strong
OH lines and conclude that the variations observed are likely from
residual sky contamination and are not significant given the signal to
noise of our data.

\section{Pure Crystalline Water Ice Model}

We produced a simple reflectance spectrum using a Hapke model and
published optical constants for crystalline water ice, similar to that
used by \cite{2005ApJ...627.1057T}.  The feature at 1.65 \micron\/,
which is obvious in our data, is found in crystalline water ice but
not amorphous water ice nor crystalline water ice at high temperatures
\citep{2004JGRE..10901012H}.  Thus, we begin our modelling of the
reflectance from 2003 $\rm EL_{61}$ with crystalline water ice only.
The wavelength dependent absorption coefficient was drawn from
\cite{1998JGR...10325809G} for 30 K, and the real index of refraction
from \cite{1984ApOpt..23.1206W} was used.  Isotropically scattering
grains were assumed, and an amplitude of opposition effect $B(0) =
2.5$ was used.  Many grain sizes were modeled: from 10 \micron\/ up to
1 cm diameter in size.  Small grain sizes of 25 \micron\/ to 50
\micron\/ were universally favored to reproduce both the depth and
width of the absorption.  We were unable to reproduce the behavior of
our spectrum shortward of 1.2 \micron\/ with any of our models.  This
region is the most uncertain in the \cite{1998JGR...10325809G}
laboratory work, with up to factors of 2 in absorption due to
fundamental limitations of their instrumental apparatus.  Such large
discrepancies can easily cause errors of $\sim 5\%$ in modelled
albedo, so we have ignored all wavelengths shortward of 1.2 \micron\/
in our fits.  The region beyond 2.35 \micron\/ is also poorly fit by
our models and was ignored in fitting.  This wavelength region is
addressed in Section~\ref{drop}.

An initial fit of the composite spectrum shows that a pure water ice
model fits the major features of the spectrum and the $J-H$ color
quite well, to a few percent (Figure~\ref{el61-waterice}).  However,
the overall $H-K$ color measured for 2003 $\rm EL_{61}$ from the
Gemini photometry, and the Keck $HK$ grism measurement is roughly 20\%
lower in $K$ flux than the pure water ice model.  Regardless of the
specific crystalline water ice temperature and grain sizes, if the
depth of the broad absorptions are matched, the overall color of the
body does not match.  It is clear that a component with a blue near-IR
color would improve the fit, but we find no evidence for specific
transitions that might indicate a particular species responsible for
the blue color.  We therefore considered several non-unique models
which could contribute to the blue color below.

\subsection{The Blue Component}

The water ice model residuals in Figure~\ref{el61-waterice} suggest
that addition of a component with an overall blue color at
near-infrared wavelengths can improve the fit substantially.  Since
there are no spectral signatures apparent besides the overall color,
it is impossible to identify the blue component accurately.  However,
a survey of published infrared absorptions shows that there are at
least a few common components that are blue in the near-infrared with
few strong absorptions.  We discuss these models below with the
understanding that it is very likely that the true composition of 2003
$\rm EL_{61}$ is quite complex and could include any, all or none of
the components listed below.  We believe it is still useful to report
on this analysis as future work may help to identify the blue
component more accurately.

The three compounds that we found to improve the fit in the
near-infrared are pure hydrogen cyanide, hydrated Titan tholin and
phyllosilicate clays.  Since the near-infrared colors were consistent
on both photometric nights at Gemini, which probed different
longitudes of the body, we believe the blue component is likely to be
present across a wide surface area of the minor planet.  A summary of
the pure water ice model and the additional blue components is
presented in Table~\ref{models} and discussion appears in following
sections.

\subsubsection{Hydrogen Cyanide}

Hydrogen cyanide (HCN) is one of the components we found to improve
the fit significantly.  A model of water ice (73\%) + hydrogen cyanide
(27\%) improves the fit significantly as estimated by $\chi^{2}$
analysis and by eye.  The best-fit water ice model in this case is a
30 K model with a 25 \micron\/ grain diameter.  The hydrogen cyanide
model used was that of room temperature HCN
\citep{2001AJ....121.1163D}.  Hydrogen cyanide is colorless in the
visible, so HCN is consistent with the neutral $V-J$ color measurement
in this work, and the neutral $V-I$ color reported by
\cite{2006retal}.  The lack of specific narrow features in HCN does
not allow unequivocal identification in our data.  Hydrogen cyanide
has several weak features at 1.52 \micron\/, 1.68 \micron\/, and 2.01
\micron\/ which are very close to nearby water absorptions and are
thus unsuitable for identification of HCN in our data.  The 1.80
\micron\/ feature is likely the most diagnostic of HCN, but it is near
two prominent OH lines, making identification of the 1.80 \micron\/
HCN feature impossible in our dataset.

\subsubsection{Hydrated Tholins}
\label{tholin}

\begin{figure}
\plotone{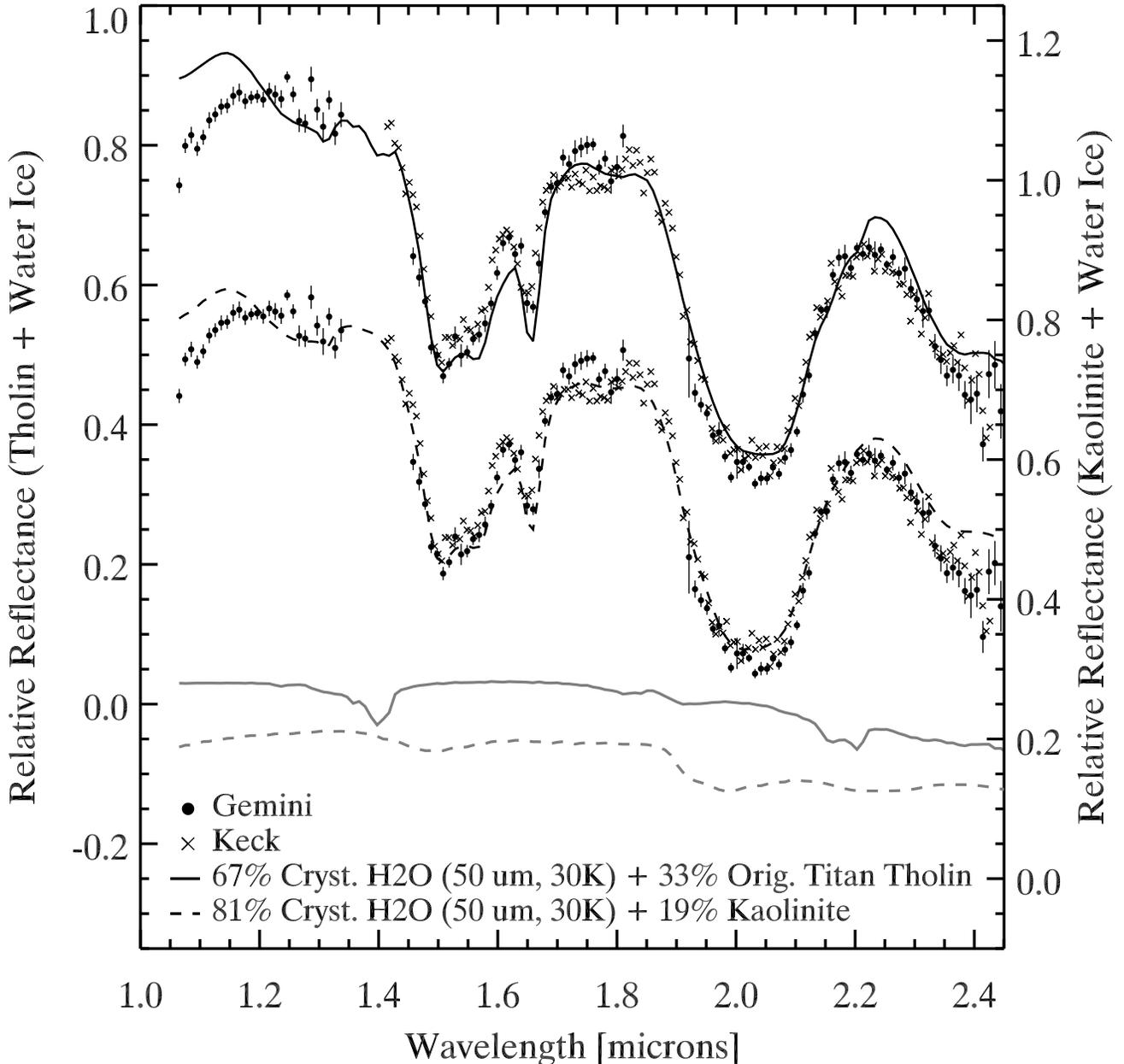} \figcaption{Relative reflectance spectrum
  of 2003 $\rm EL_{61}$ from Gemini and Keck (black filled circles and
  x's) normalized to the model.  Overplotted is our best fit
  crystalline water ice + hydrated tholin model (solid black line) at
  its true geometric albedo (left vertical axis).  Also is our best
  fit crystalline water ice + kaolinite clay model (dashed black line)
  at its true geometric albedo (right vertical axis).  The individual
  tholin (solid gray, offset downward by -0.12 for clarity) and
  kaolinite components (dashed gray) are plotted.  Note that although
  tholins improve the fit in the near infrared, such large amounts are
  ruled out by the neutral visible color of 2003 $\rm EL_{61}$.  See
  text for full details of the models.
\label{el61-tholinclay}}
\end{figure}

The effect of adding hydrated tholins to the fit was estimated by using
the 3 tholin types discussed by \cite{2004Icar..168..158R}.  The three
types were the original Titan tholin manufactured by
\cite{1984Icar...60..127K} and measured by \cite{1991Icar...94..345C},
as well as the Denver and Ames tholins \citep{2004Icar..168..158R}.
Addition of all three tholins improved the fit, but only the original
Titan tholin, which was less hydrated than the other samples, improved
the fit at the statistically significant ($> 3\sigma$) level as judged
by a $\chi^2$ analysis.  The best results occurred for original tholin
with 67\% water ice at 30 K with 50 \micron\/ grains + 33\% original
tholin, shown in Figure~\ref{el61-tholinclay}.  We remark that there
are a wide variety of Titan tholins published, and even multiple
measurements of apparently identical compounds also have resulted in
discrepant laboratory spectra, adding uncertainty to all tholin
models.  The strongest evidence against tholins comes from our $V-J$
color, which is neutral in reflectance and the neutral $V-I$ color
reported by \cite{2006retal}.  The addition of substantial amounts
(i.e. the 1/3 in our best-fit spectrum) of tholin material would
result in a very red surface, so although the original Titan tholin
fits the near-infrared spectrum, we consider tholins unlikely to
reproduce the visible spectrum of 2003 $\rm EL_{61}$, which is
colorless throughout the visible.

\subsubsection{Phyllosilicate Clays}
\label{clays}

The addition of the common Earth phyllosilicate clay kaolinite
\citep[W1R1Bb sample]{splib} also improves the fit somewhat.  A
$\chi^{2}$ analysis shows that the amount of improvement falls short
of formal significance, yet it was a larger improvement than for any
other component tested and produced an improvement to the eye, so we
include it here.  The best-fit model is composed of 81\% crystalline
water ice with 50 \micron\/ grains at 30K + 19\% kaolinite, shown in
Figure~\ref{el61-tholinclay}.  The common phyllosilicate clay
montmorillonite \cite{2005Icar..179..259R} was also found to improve
the fit, but to even a smaller degree than kaolinite.  Since the
presence of phyllosilicate clays improves the overall fit of the
spectrum, we consider these to be a possible source of the
near-infrared blue color of 2003 $\rm EL_{61}$.  Most montmorillonite
and kaolinite clays are colorless or nearly so in the visible, so we
consider these to be a reasonable component to reproduce the blue
color of the near-infrared spectrum without affecting the visible
color of 2003 $\rm EL_{61}$.

\subsubsection{Failed Models}

We tested many other models that are somewhat blue in the near
infrared as shown in Table~\ref{models} and the references therein.
Several models provided no improvement at all over the water ice model
including methanol ice \citep{1998Icar..135..389C}, asphaltite and
kerite \citep{2004Icar..170..214M}.  Several models showed some
improvement, but not at the formally significant level: ammonia
hydrate \citep{2001AJ....121.1163D}, ammonia ice
\citep{2000Sci...287..107B}, amorphous water ice \citep[best fit with
25 \micron\/ grains]{1998ssi..conf..199S}, methane ice \citep[best fit
with 50 \micron\/ grains]{2002Icar..155..486G}, montmorillonite
\citep{2005Icar..179..259R} and the Ames and Denver tholin compounds
\citep{2004Icar..168..158R}.  A few of these compounds deserve
additional comment.  Tholin compounds and montmorillonite were
discussed above in Sections~\ref{tholin} and \ref{clays},
respectively.  The inclusion of asphaltite and kerite are mentioned
below, in Section~\ref{drop}.

Ammonia ice and ammonia hydrate have been reported on Charon and on
Quaoar \citep{2000Sci...287..107B,2004Natur.432..731J} and are of
particular interest because when dissolved in water, they lower its
melting point.  Such compounds may make it easier to resurface a body
with crystalline water ice, but we find no evidence of such compounds
in our data.

Amorphous water is known to co-exist with crystalline water ice on the
Galilean satellites \citep{2004JGRE..10901012H}, thus it seems
possible that this could be the case for 2003 $\rm EL_{61}$, although
we find no formal evidence for this in our data.  It is possible that
large amounts of the water ice could be amorphous (up to $\sim 1/3$)
without detection in our data because of the large degeneracy between
crystalline water ice grain size, temperature, and amount of amorphous
water ice, all of which affect the depth of the 1.65 \micron\/
crystalline water ice feature.  Much higher signal to noise spectra in
the $H$ band would place a stronger limit on the amount of amorphous
water ice.

Although the addition of methane ice does not significantly improve
our fit, because it is a known component on other solar system bodies
and highly volatile, it is interesting to ask how much could be on the
surface without being detected.  We find that the addition of 10\%
methane ice (grain size of 50 \micron\/) results in a significant ($>3
\sigma$) deviation of the model from the data in terms of $\chi^2$ as
well as by eye.

\subsection{Neutral Absorber}

We also modeled the presence of a neutral absorber; the addition of a
neutral absorber of any albedo did not improve the overall fit
significantly for the water ice + blue component models.  Our
reflectance models alone cannot specifically test for the presence of
highly absorbing, featureless compounds such as carbon black.  The
addition of such an absorber would not change the relative reflectance
or the depths of the absorption bands we observe, only the overall
albedo of the body.  However, by using published albedo measurements,
we can place some constraint on the presence of dark absorbers.

\begin{figure}
\plotone{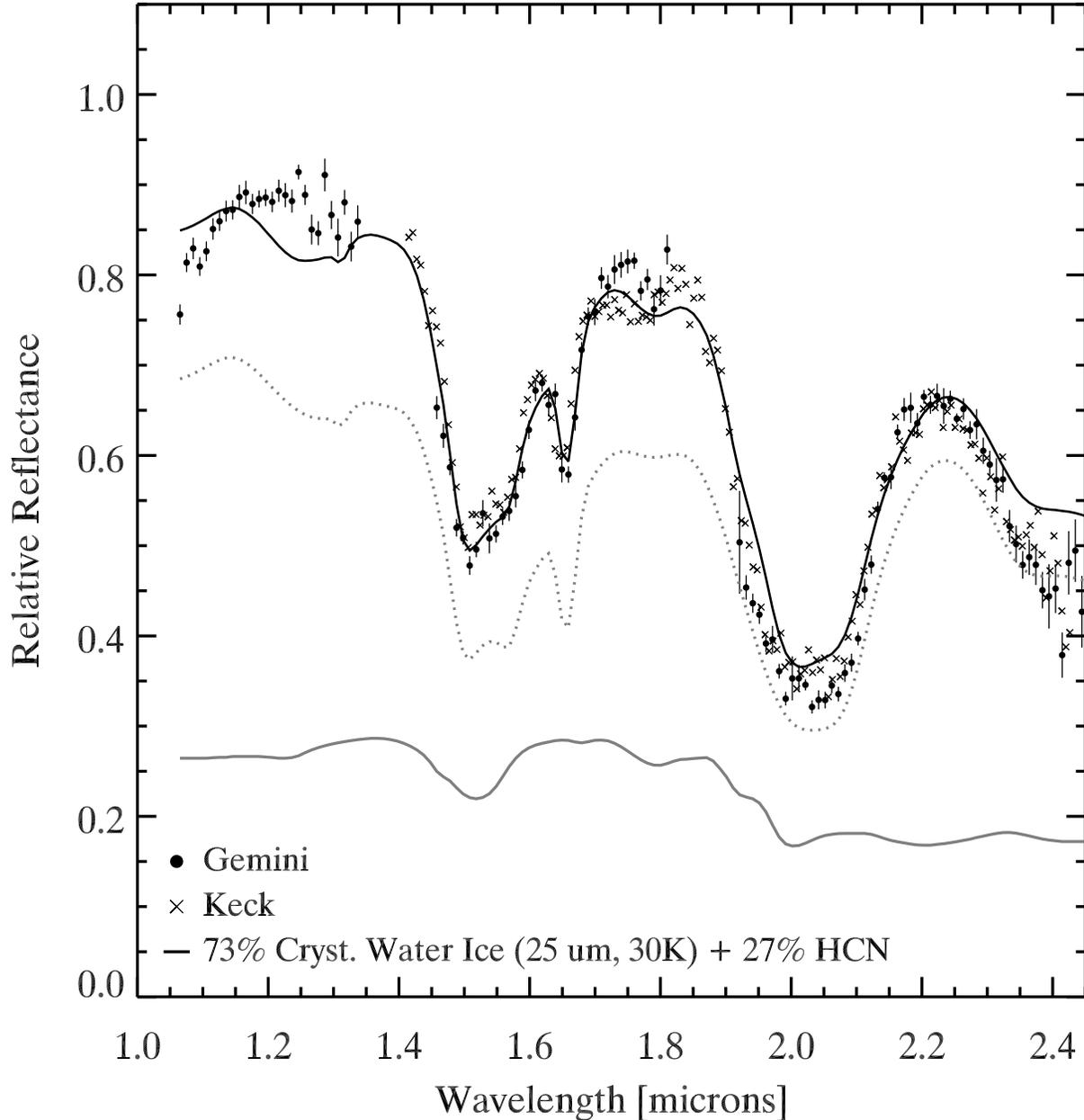} \figcaption{Relative reflectance spectrum of
  2003 $\rm EL_{61}$ from Gemini and Keck (black filled circles and
  x's) normalized to the model.  Overplotted is our best fit pure
  crystalline water ice model with cyanide (HCN) at its true geometric
  albedo.  The individual water ice and HCN components (shifted
  vertically by 0.1 for clarity) appear as a dotted line and solid
  gray line respectively.  See text for full details of the models.
\label{el61-hcn}}
\end{figure}

Previously, \cite{2006retal} placed constraints on the albedo of 2003
$\rm EL_{61}$ from lightcurve and binary orbit measurements.  The
spheroidal and ellipsoidal models place the $V$-band albedo in the
$0.6 < p_V < 0.8$ range.  Such albedo measurements are consistent with
our models, which include no dark or neutral materials.  Since the
$V-J$ color is roughly solar, then the $J$ albedo and $V$ albedo are
roughly equal.  Using this information, we find that all of the
two-component models in Figures~\ref{el61-hcn}--\ref{el61-tholinclay} fall
within $0.75 < p_V < 0.90$, showing a reasonable consistency between
the dynamical spin models and our icy near-infrared reflectance
models.  Thus, we expect the true surface fractions of ices on 2003
$\rm EL_{61}$ to be very close to the reflectance model fractions
presented here, with very little low albedo material present on the
surface.

\subsection{The 2.35 \micron\/ Drop}
\label{drop}

We note that there is a clear drop in the spectrum of 2003 $\rm
EL_{61}$ beyond 2.35 \micron\/ compared to the crystalline water ice
model as well as the water ice + blue component models.  This drop
is observed in both the Keck spectrum and the Gemini spectrum.  The
feature appears to start at about 2.35 \micron\/ and continues until
atmospheric sky noise overcomes the object signal at $\sim 2.45$
\micron\/, indicating that the central wavelength of this absorption
is at or beyond 2.45 \micron\/.  High signal to noise observations of
Charon have shown a similar feature, but the carrier was never
identified \citep{2000Sci...287..107B}.  A similar absorption has been
reported on Phoebe, and has been attributed to a cyanide combination,
possibly potassium cyanide or cyanide trihydrate
\citep{2005Natur.435...66C}.  Such absorption has also been reported
by \cite{2005AGUFM.P22A..02C} for Iapetus, Dione, and the F-ring.
Judging from \cite{2005Natur.435...66C}, the reflectance of 2003 $\rm
EL_{61}$ appears to be most consistent with copper potassium cyanide
in this region, as it is broad ($\sim 0.1$ \micron\/) and in the
correct location.  We did not specifically model this compound in more
detail because we were unable to find data on its absorption shortward
of 1.8 \micron\/ in any available literature or chemical database.
Although potassium cyanide and cyanide trihydrate may be responsible
for the observed drop, other inorganic cyanide salts ($\rm \sbond
C \tbond N$) are not a good match.  Nickel cyanide, cadmium cyanide
and zinc cyanide all produce very narrow transitions (widths about
0.03 \micron\/) shortward of 2.36 \micron\/ \citep{splib}.  Organic
nitrile compounds ($\rm \sbond C \sbond N \tbond C$) share the same
$\rm C \tbond N$ triple bond of inorganic cyanide salts, but we were
unable to find suitable spectral information for these components.
The addition of bitumens such as kerite, asphaltite or wurtzilite,
which have a drop at 2.3 \micron\/ could also explain the 2.35
\micron\/ drop \citep{1998Icar..134..253M}.  We examined the use
of both powdered kerite and asphaltite \citep{2004Icar..170..214M},
but neither appeared to improve the overall fit significantly due to
the steep red slope seen at wavelengths shorter than 1.5 \micron\/.

\section{Discussion}
\label{discussion}

The basic result of our observations is that the spectrum of 2003 $\rm
EL_{61}$ is consistent with a model with roughly 2/3 to 4/5 pure
crystalline water ice combined with 1/3 to 1/5 pure blue material.
The best candidates for the blue material are pure hydrogen cyanide or
possibly phyllosilicate clays, although this identification is not
unique.  Likely there are many possible materials that could fit the
near-infrared color of 2003 $\rm EL_{61}$, of which only a fraction
have laboratory spectra available.  No neutral absorbers are needed in
appreciable amounts to model the relative reflectance of the body.
The only feature noted in the near infrared spectrum besides that of
crystalline water ice and the basic blue color is the drop in
reflectance beyond 2.3 \micron\/ which is consistent with copper
potassium cyanide.  Laboratory data on cyanide compounds are sparse,
so it is likely that other cyanide or nitrile compounds could produce
such behavior.

The presence of crystalline water ice has been widely reported for
many outer solar system bodies such as the satellites of the major
planets, Charon and most recently Quaoar.  As noted by
\cite{2004Natur.432..731J} for (50000) Quaoar, the presence of
crystalline water ice is considered problematic in the Kuiper belt
because, in general, lower energy amorphous ice is favored in cold
environments.  Crystalline water ice may have been prevalent in the
early solar nebula and may have been incorporated into bodies at low
temperatures ($< 100$ K) if the deposition flux of water molecules was
low \citep{1994A&A...290.1009K}.  However, the disruptive action of
cosmic ray and solar flux bombardment on the outer surfaces of icy
bodies is expected to rearrange crystalline water ice to amorphous in
the outer solar system \citep{2004JGRE..10901012H}.

Both solar flux and cosmic ray bombardment can cause extensive
radiation damage to the outer surface layers of KBOs.  Estimating the
time scale of such radiation damage is very difficult due to the
sparse amount of data available about energetic particles in the solar
system, which comes primarily from spacecraft.  In addition,
assumptions about the timescale for density changes in the local
interstellar medium must be made which directly affects all estimates
of the extent of the heliosphere, which is a key component of
radiation damage models.  Our near-infrared reflectance spectra probe
only the first $\sim 1$ mm of the KBO surface when surface density and
multiple scattering effects are taken into account.
\cite{2003EM&P...92..261C} suggest that disruptive amounts of energy
(of order $\sim 100$ eV per 16 amu) can be delivered to $\sim 1$ mm
depths by Galactic Cosmic Ray bombardment throughout the heliosphere.
Timescales several orders of magnitude shorter may be plausible both
closer to and farther from the sun than the $\sim 40$ AU region of the
Kuiper Belt due to proton bombardment from a variety of sources.  As
many of these KBOs have been at their present location for $10^9$
years even in extreme dynamical models \citep{2005Natur.435..466G}, it
appears that a solar system wide model for surface renewal is likely
needed to explain the presence of crystalline water ice.

Hydrogen cyanide, one of the species that can explain the blue near
infrared color of 2003 $\rm EL_{61}$, is also photolysed rapidly in
the outer solar system environment.  Whether alone or frozen in
solution with water ice, hydrogen cyanide is destroyed after
deposition of about 100 eV per molecule by both irradiation and
photolysis \citep{2005ApJ...620.1140G}.  Thus if HCN is to be
considered as a source of the blue material on 2003 $\rm EL_{61}$, it
has the same stability problems as crystalline water ice.  The
phyllosilicate clays, another possible infrared blue material, do not
share this sensitivity to radiation damage.

\section{Summary}

Using the Gemini and Keck telescopes, we have measured the
near-infrared reflectance spectrum of 2003 $\rm EL_{61}$ on a global
scale.  We find the following:

\begin{itemize}

\item The presence of crystalline water ice dominates the spectrum.
  Best fit models require between 2/3 and 4/5 of the reflectance to be
  attributed to pure crystalline water ice.

\item As for other outer solar system bodies, crystalline water ice is
  unstable on solar system timescales, suggesting some kind of global
  surface renewal process.

\item The addition of a second component that is blue in the
  near-infrared dramatically increases the goodness of fit and is
  responsible for 1/3 to 1/4 of the reflectance.  The most likely
  carriers for this component are hydrogen cyanide or phyllosilicate
  clays, although this identification is not unique.  Although
  hydrated titan tholins do improve the near-infrared fit, they are
  inconsistent with the neutral visible color observed for 2003 $\rm
  EL_{61}$.

\item The presence of inorganic cyanide salts such as copper potassium
  cyanide may explain the observed absorption beyond 2.3 \micron\/, as
  has been postulated for Phoebe and other Saturnian bodies
  \citep{2005Natur.435...66C}.

\item There is likely to be very little low albedo material on the
  surface of 2003 $\rm EL_{61}$ as our pure ice model albedos are
  consistent with published lightcurve/orbit dynamical constraints
  \citep{2006retal}.  Thus, the fractions described for reflectance
  are similar to the global surface coverage for the ices described.
  This surface coverage can only be estimated to the $\sim 1$ mm
  depth, due to our wavelength of observation.

\item No more than $\sim 10$\% of the surface of 2003 $\rm EL_{61}$
  could be covered with methane ice without detection.

\item The near-infrared photometric differences between the two epochs
  at which we observed are consistent with the previously published
  visible lightcurve of 2003 $\rm EL_{61}$, which is attributed to
  pronounced global rotational distortion of the body.  The ubiquity
  of these variations from 0.4 -- 2.5 \micron\/ adds more evidence to
  the conclusion that the extreme lightcurve of 2003 $\rm EL_{61}$ is
  due to shape only.

\end{itemize}

\acknowledgements

We appreciate the comments of the anonymous referee who reviewed this
work.  We thank the Gemini science staff who scheduled our program,
operated the telescope and collected the queue mode data: Tracy Beck,
Julia Bodnarik, Simon Chan, Avi Fhima, Rachael Johnson, Inger J\o
rgensen, Andrew Stephens, Kevin Volk, and Dolores Walther.  We thank
the observing assistants and support astronomers at the W. M. Keck
Observatory who helped with this project: Marc Kassis, Chuck Sorenson
and Steven Magee.  This work was based on observations obtained at the
Gemini Observatory, which is operated by the Association of
Universities for Research in Astronomy, Inc., under a cooperative
agreement with the NSF on behalf of the Gemini partnership: the
National Science Foundation (United States), the Particle Physics and
Astronomy Research Council (United Kingdom), the National Research
Council (Canada), CONICYT (Chile), the Australian Research Council
(Australia), CNPq (Brazil) and CONICET (Argentina).  Observations were
collected under Gemini program IDs GN-2004B-Q-56 and GN-2005A-Q-45.
Some of the data presented herein were obtained at the W. M. Keck
Observatory, which is operated as a scientific partnership among the
California Institute of Technology, the University of California, and
the National Aeronautics and Space Administration. The Observatory was
made possible by the generous financial support of the W. M. Keck
Foundation.

\begin{deluxetable}{rrrrrr}
\rotate
\tablecolumns{6}
\tablecaption{Gemini Observations of 2003 $\rm EL_{61}$}
\tablehead{
\colhead{UT date} & \colhead{UT time} & \colhead{Grating} & \colhead{Airmass} & \colhead{Exposure}   & \colhead{Sky} \\
\colhead{}        & \colhead{}        & \colhead{}        & \colhead{}        & \colhead{Time [min]} & \colhead{} }   
\startdata
2005 Jan 27 & 13:36 -- 14:46 & $K$-grism      & 1.098 -- 1.013 & 65 & Thin Cirrus \\
2005 Jan 27 & 15:51 -- 16:18 & $K$-grism      & 1.008 -- 1.024 & 25 & Photometric \\
2005 Mar 02 & 10:34 -- 10:40 & $J$ photometry & 1.239 -- 1.218 &  4 & Photometric \\
2005 Mar 02 & 10:40 -- 10:47 & $H$ photometry & 1.215 -- 1.196 &  4 & Photometric \\
2005 Mar 02 & 10:47 -- 11:08 & $K$ photometry & 1.193 -- 1.135 &  4 & Photometric \\
2005 Mar 02 & 11:25 -- 12:18 & $K$-grism      & 1.091 -- 1.024 & 50 & Photometric \\
2005 Mar 02 & 12:24 -- 13:17 & $H$-grism      & 1.015 -- 1.000 & 50 & Photometric \\
2005 Mar 02 & 13:21 -- 13:53 & $J$-grism      & 1.002 -- 1.015 & 30 & Photometric \\
2005 Apr 25 & 07:25 -- 07:31 & $J$ photometry & 1.153 -- 1.138 &  4 & Photometric \\
2005 Apr 25 & 07:32 -- 07:38 & $H$ photometry & 1.136 -- 1.122 &  4 & Photometric \\
2005 Apr 25 & 07:38 -- 07:44 & $K$ photometry & 1.120 -- 1.107 &  4 & Photometric \\
2005 Apr 25 & 08:04 -- 08:57 & $K$-grism      & 1.065 -- 1.013 & 50 & Photometric \\
2005 Apr 25 & 11:29 -- 12:09 & $J$-grism      & 1.133 -- 1.239 & 25 & Photometric \\
2005 Apr 25 & 12:10 -- 12:41 & $H$-grism      & 1.260 -- 1.381 & 30 & Photometric \\
2005 May 31 & 08:56 -- 09:59 & $K$-grism      & 1.111 -- 1.294 & 50 & Cirrus      \\
2005 May 31 & 10:00 -- 11:01 & $J$-grism      & 1.320 -- 1.688 & 45 & Cirrus      \\
\enddata
\label{observations}
\end{deluxetable}

\begin{deluxetable}{rrlrrr}
\tablewidth{7in}
\tablecolumns{6}
\tablecaption{Spectral Models of 2003 $\rm EL_{61}$}
\tablehead{
\colhead{Water Grain}    & \colhead{Water Ice} & \colhead{Secondary} & \colhead{Secondary} & \colhead{Improvement over} & \colhead{Visible} \\
\colhead{Size [\micron\/]} & \colhead{Fraction}  & \colhead{Component} & \colhead{Fraction} & \colhead{water ice model} & \colhead{Color} }
\startdata
 25 & 100\% & 50 \micron\/ Water ice\tablenotemark{a} &  0\% &          N/A  & Neutral  \\
 25 &  73\% & HCN\tablenotemark{b} & 27\% & Significant   & Neutral \\
 50 &  67\% & Original tholin\tablenotemark{c} & 33\% & Significant   & Red     \\
 50 &  81\% & Kaolinite\tablenotemark{d}       & 19\% & Borderline    & Neutral \\
    &       & Ammonia hydrate\tablenotemark{b} &      & Insignificant & Neutral \\
    &       & Ammonia ice\tablenotemark{e}     &      & Insignificant & Neutral \\
  & & 25 \micron\/ Amorphous Water\tablenotemark{f} & & Insignificant & Neutral \\
  & & 50 \micron\/ Methane ice\tablenotemark{g}     & & Insignificant & Neutral \\
    &       & Montmorillonite\tablenotemark{h} &      & Insignificant & Neutral \\
    &       & Ames tholin\tablenotemark{c}     &      & Insignificant & Red     \\
    &       & Denver tholin\tablenotemark{c}   &      & Insignificant & Red     \\
    &       & Methanol ice\tablenotemark{i}    &      & None          & Neutral \\
    &       & Asphaltite\tablenotemark{j}      &      & None          & Red     \\
    &       & Kerite\tablenotemark{j}          &      & None          & Red     \\
\enddata
\tablenotetext{a}{\cite{1998JGR...10325809G} and \cite{1984ApOpt..23.1206W}}
\tablenotetext{b}{\cite{2001AJ....121.1163D}}
\tablenotetext{c}{\cite{2004Icar..168..158R}}
\tablenotetext{d}{\cite{splib}}
\tablenotetext{e}{\cite{2000Sci...287..107B}}
\tablenotetext{f}{\cite{1998ssi..conf..199S}}
\tablenotetext{g}{\cite{2002Icar..155..486G}}
\tablenotetext{h}{\cite{2005Icar..179..259R}}
\tablenotetext{i}{\cite{1998Icar..135..389C}}
\tablenotetext{j}{\cite{2004Icar..170..214M}}
\tablecomments{Compounds that are neutral or nearly so
  in the visible are favored because of the neutral visible color of
  2003 $\rm EL_{61}$.  See text for complete discussion.}
\label{models}
\end{deluxetable}

\end{document}